\documentclass[12pt]{article}
\usepackage{authblk}
\usepackage[top=0.6in,bottom=0.4in,left=0.7in,right=0.2in]{geometry}
\usepackage{amssymb}
\usepackage{graphicx}
\usepackage{hyperref}
\usepackage{caption}
\newcommand{\chb}{\textcolor{blue}}

\hypersetup{
               colorlinks=true,
               linkcolor=blue,
	       citecolor=blue,
               filecolor=magenta,
               urlcolor=cyan,
}
\begin{document}
\title{The effect of antisite disorder on magnetic and exchange bias properties of Gd-substituted Y$_2$CoMnO$_6$ double perovskite. }

\author[1*]{Anasua Khan}
\author[2]{Sarita Rajput}
\author[2]{M. Anas}
\author[2]{V. K. Malik}
\author[2]{T. Maitra}
\author[1]{T.K Nath}
\author[1]{A.Taraphder}

\affil[1]{Department of Physics, Indian Institute of Technology, Kharagpur-721302, India}
\affil[2]{Department of Physics, Indian Institute of Technology Roorkee, Roorkee-247667, India}
\affil[*]{anasua.khan@iitkgp.ac.in}

\setcounter{Maxaffil}{0}
\renewcommand\Affilfont{\itshape\small}
\date{\today}

\maketitle
\begin{abstract}

Combining experimental investigations and first-principles DFT calculations, we report physical and magnetic properties of Gd-substituted Y$_2$CoMnO$_6$ double perovskite, which are strongly influenced by antisite-disorder-driven spin configurations. On Gd doping, Co and Mn ions are present in mixed-valence (Co$^{3+}$, Co$^{2+}$, Mn$^{3+}$ and Mn$^{4+}$) states. Multiple magnetic transitions have been observed: i) paramagnetic to ferromagnetic transition is found to occur at \textit{T}$_C$=95.5 K, ii) antiferromagnetic transition at \textit{T}$_N$=47 K is driven by $3d-4f$ polarisation and antisite disorder present in the sample, iii) change in magnetization below \textit{T}$\leq$20 K, primarily originating from Gd ordering, as revealed from our DFT calculations. AC susceptibility measurement confirms the absence of any spin-glass or cluster-glass phases in this material. A significantly large exchange bias effect (\textit{H}$_{EB}$=1.07 kOe) is found to occur below 47 K due to interfaces of FM and AFM clusters created by antisite-disorder.

\end{abstract}

\section{Introduction}

Multifunctional insulators with strong magnetoelectric coupling, room-temperature ferromagnetism and colossal magnetoresistance have sparked substantial research interests during the last two decades \cite{int_2}. In this regard, double perovskite (DP) materials with the general formula R$_ 2$BB$^{\prime}$O$_6$ (R= Y, rare earth, B/B$^{\prime}$ transition metal) show many promises: they exhibit intriguing properties like magnetodielectric, magnetocapacitance, exchange bias anisotropy, magnetocaloric effect\cite{int_4, int_6, int_7}, and others, which account for their wide range of applications in spintronics, sensors, tunable microwave filters, energy harvesting devices and transformers\cite{int_9}. The strong interplay between spin, charge, and lattice degree of freedom is responsible for such interesting multifunctional characteristics\cite{int_12,int_13}.

In the DP family, La$_2$CoMnO$_6$ (LCMO) and La$_2$MnNiO$_6$ (LNMO) have been widely investigated. Several intriguing properties in their ordered and disordered phases have been reported\cite{int_14,int_15}. There have been fewer investigations on double perovskites that include rare earth having smaller cationic radii\cite{int_16}. The rare earth cationic size in the DP system plays a vital role in their physical and magnetic properties. The decrease in rare earth cationic size significantly changes the B-O-B$^{\prime}$s bond angle and bond length (B-O, B$^{\prime}$-O). As a result, frustration occurs in the structure associated with an octahedral tilt which affects the magnetic and electronic properties\cite{int_17}. For example, the A-type AFM magnetic ground state of RMnO$_3$ gets transformed into an E-type AFM ground state depending upon the rare earth ion size\cite{int_18,int_19}. Furthermore, the RCo$_{0.5}$Mn$_{0.5}$O$_3$ (R=Eu, Tb, Y) system exhibits metastable behavior in the presence of an external magnetic field, although the spontaneous magnetization is lower than that of LCMO\cite{int_20,int_22}. On the other hand, the introduction of Y/Yb/Lu at La site (R$_2$CoMnO$_6$) will induce ($\uparrow \uparrow \downarrow \downarrow$) E-type AFM ordering, which breaks the inversion symmetry and gives rise to electric polarisation\cite{int_23,int_24,int_25}. At the same time, the other rare earth materials like Ho, Tm, and Er, having smaller ionic radii, show usual ferromagnetism\cite{int_26}. In particular, the Y-based DP compounds have drawn significant interest due to their magnetic ground states. The magnetic ground state of Y$_2$CoMnO$_6$ (YCMO) is still under debate. The first-principle calculations reveal that the ground state comprises E-type AFM in competition with FM and A-type magnetic ordering\cite{int_27}. However, E-type order is not observed experimentally\cite{int_28}.

Interestingly, the structural, magnetic and electronic properties of DPs can be tuned when the A-site element is partially substituted by cations with different ionic radii. For example, when the Eu$^{3+}$ ion is partly replaced with the Y$^{3+}$ ion in EuMnO$_3$, structural distortion affects the magnetic and multiferroic properties\cite{int_29,int_30}. The mismatch in ionic radii at the A-site induces quench disorder and local distortion in the structure. Therefore, the frustration in the structure or the random Columb potential, due to different valence states\cite{int_31}, works as a stimulating agent and suppresses ordering parameters like magnetism, charge order and superconductivity\cite{int_32,int_33,int_34}. However, interesting phenomena such as multiglass behavior, phase separation, exchange bias and ferroelectricity\cite{int_35,int_36,int_37} also appear out of such frustration. Indeed, the properties become more interesting when the A site dopant is a magnetic rare-earth. An interesting problem is to unravel such possibilities in YCMO and to this end, we have performed experimental as well as the first-principles study of partially Gd-substituted YCMO. Notably, we focus on the case of YGdCoMnO$_6$, where 50\% Y$^{3+}$ (r=1.075 \AA) is replaced by Gd$^{3+}$ (r=1.27 \AA) ion. The rare earth Gd gains more attention because of its significant magnetic moment and room temperature magnetocaloric properties\cite{int_38}. Also, the 3d-4f exchange interaction (originating from interactions between the 3d orbital of TM and the complex, localized 4f orbital of Gd) has a more significant impact on changing the ground-state magnetic order. The magnetic nature of the A-site ion produces different magnetic interactions in competition with the 3d magnetic ion and makes the system highly frustrated, resulting in different magnetically ordered phases.

Ferromagnetism in the DP systems is due to superexchange interaction between an ordered arrangement of B$^{2+}$ and B$^{\prime}$$^{4+}$ cations. In a pristine sample, however, a perfectly ordered arrangement is impossible to accomplish. Antisite disorder (ASD) (an imperfection caused by the misplacement of BO$_6$ and B$^{\prime}$O$_6$ octahedra) often occurs during sample synthesis. As a result, more B-O-B and B$^{\prime}$-O-B$^{\prime}$ couplings are formed, increasing the lattice strain and causing frustration in the structure. Moreover, strong structural distortion indicates competitive magnetic interactions between short-range FM next-nearest-neighbor interaction and long-range AFM nearest-neighbor interaction\cite{int_39}. Therefore, ASD breaks the long-range ferromagnetic ordering and drives the system towards a magnetic glassy order\cite{int_40}. Murthy \textit{et al.} show that the coupling between FM and AFM phases increases with Sr doping at La site in LCMO\cite{int_41}. Again, first-principle calculations show that the ASD increases with Sr doping at La site in LCMO\cite{int_9}. The present work deals with the occurrence of ASD with Gd doping at the Y site and its effects on the magnetic properties of the system.

The Exchange bias (EB) phenomenon is generally observed in the magnetically phase-separated system. The anisotropy arises due to exchange interaction at the interface between the different magnetic ordered phases giving rise to EB effect\cite{int_48}. The interfaces can be ferromagnetic/antiferromagnetic, ferromagnetic/ferrimagnetic, antiferromagnetic/spin glass\cite{EX_6}, and so on. Materials with large EB and coercivity are attracting a lot of attention these days due to more significant impact on technology owing to its application in spin-valves\cite{int_42}, read-heads in magnetic recording\cite{int_43}, permanent magnets\cite{int_44}, giant magnetoresistive random access memory\cite{int_45} and other devices. Moreover, the EB effect is observed in many phase-segregated systems (where phase segregation occurs spontaneously). Generally, in that case, the compounds belong to a single phase at high temperature, and multiple magnetic phases are originated with the lowering of temperature. Such kind of feature was observed in Pr$_{2/3}$Ca$_{1/3}$MnO$_3$\cite{int_46} and Nd$_{2/3}$Ca$_{1/3}$MnO$_3$\cite{int_47}. We discuss here the correlation between observed field-induced EB effect and complex magnetization in Gd-substituted YCMO DP system.


\section{Experimental and Computational Details}

Using the standard Sol-gel method, polycrystalline, single-phase double perovskite compound of Y$_{2-x}$Gd$_x$CoMnO$_6$ ($x$=1.0) has been synthesized. In starting, the compounds Y$_2$O$_3$, Gd$_2$O$_3$, Co(No$_3$)$_2$, 6H$_2$O and Mn(CH$_3$COO)$_2$,4H$_2$O were taken. A stoichiometric amount of rare earth oxides (Y$_2$O$_3$, Gd$_2$O$_3$) were dissolved in dilute nitric acid to convert them into nitrate form and the rest of the precursors were dissolved in deionized water. After that, two solutions were mixed with the addition of citric acid as a chelating agent. The resulting solution was first heated at 90 $^{\circ}$C for 2 h then finally heated at 150 $^{\circ}$C for 12 h to evaporate the solvents entirely. The obtained black fluffy powder was calcined at 1100 $^{\circ}$C for 10 h. To make a pellet of 10mm diameter for measuring purpose, this powder was employed using hydraulic press and sintered at 1200$^{\circ}$C for 12 h.  

X-ray diffraction (XRD) measurement was carried out to investigate the crystal structure of bulk YGdCMO sample using X'pert PRO from Philips Pananalytical HRXRD-I PW 3040/60 of Cu-K$_{\alpha}$ radiation ($\lambda$= 1.5406 \r{A}) at room temperature. The Rietveld refinement\cite{EXP-1} of the recorded XRD pattern was executed using FULLPROF software\cite{EXP-2}. The electronic charge state of Co, Mn, and oxygen in the sample was determined using X-ray photoelectron spectroscopy (XPS, PHI 5000 versa Probe II scanning) at room temperature. The field cool (FC) and zero fields cool (ZFC) dc magnetization measurements have been performed in the temperature ranging from 5 - 300K. This was performed under the application of various dc magnetic fields by using a Quantum Design SQUID magnetometer. AC susceptibility measurements have been carried out by using PPMS (Physical property measurement system).

Density functional theory (DFT) calculations are performed to obtain electronic structure of Gd-substituted Y$_2$CoMnO$_6$ using pseudo-potential and plane-wave based method as implemented in the Vienna \textit{ab initio} simulation program (VASP)\cite{vasp-1}. In our calculations we have used Perdew-Burke-Ernzerhof generalized gradient approximation (PBE-GGA)\cite{vasp-2} and included Coulomb correlation within GGA+U approximation\cite{vasp-3}. Structural parameters were obtained from the experiment and relaxed keeping the Gd 4f states frozen as in the core. Ionic positions were relaxed until the forces on the ions are less than 0.1 meV. We have taken the supercell in \textit{c}- direction to implement the calculation. To find out the magnetic ground structure of the system, we have performed various magnetic calculations considering ferromagnetic and antiferromagnetic arrangements among the spins of Gd, Mn and Co ions. For the subsequent self-consistent calculations, the Gd 4f states were treated as valence states. An energy cut-off of 500 eV was used for the plane waves in the basis set, while a 6×6×6 Monkhorst-Pack k-mesh centered at $\Gamma$ was used for performing the Brillouin zone integrations.


\section{Results and discussion}
\subsection{Crystal structure}
\begin{figure*}[htp]
\centering
\includegraphics[scale=.4]{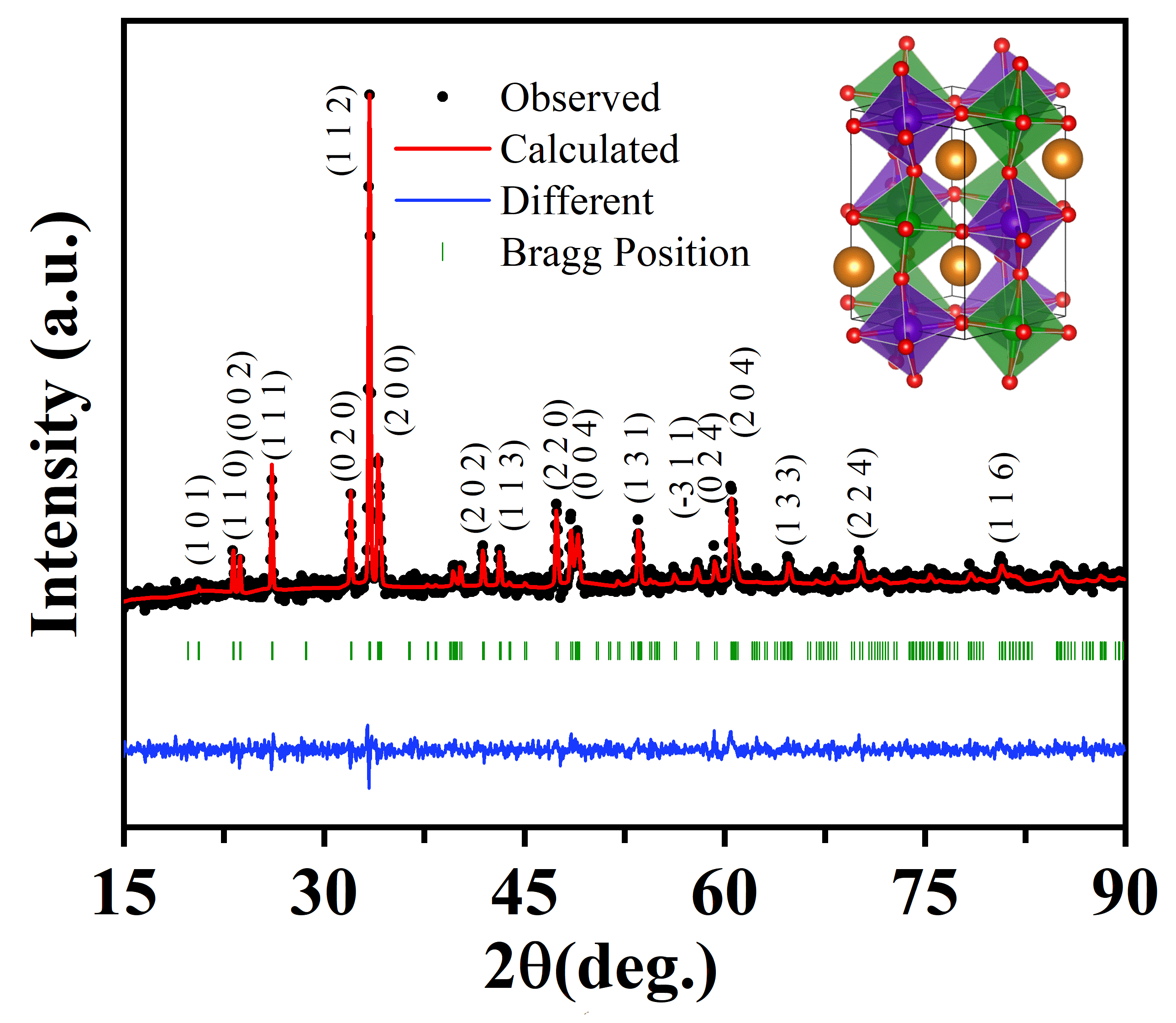}  
\caption{X-ray powder diffraction pattern along with Rietveld refinement of YGdCMO. The inset represents a schematic of a unit cell of monoclinic YGdCMO, where large golden spheres represent Y/Gd atoms sitting in the hollow formed by CoO$_6$(green) and MnO$_6$(purple) octahedra. Small red spheres represent oxygen atoms. }
\end{figure*}
In the present experiment, the XRD technique was used to determine the phase purity of crystal structure, lattice parameter, bond length, and bond angle of powder YGdCMO. Figure 1 shows the recorded XRD data along with Reitveld refinement at room temperature. The YGdCMO is found to crystallize in monoclinic symmetry of space group $P2_1/n$ as a major phase and with orthorhombic symmetry of space group \textit{Pnma} as a minor phase. The major reflection peaks are found to fit well with monoclinic symmetry rather than orthorhombic symmetry. The concentration of the monoclinic phase is found to be 99.42\%, whereas 0.58\% for the orthorhombic phase. As the dominatnt phase is monoclinic, we continue our discussion based on this phase.

The monoclinic space group is commonly known as a distorted space group. A schematic of the crystal structure of monoclinic YGdCMO has been shown in the inset of Figure 1. Six oxygen atoms octahedrally surround Co and Mn atoms. Here, every CoO$_6$ octahedra share its corner with six MnO$_6$ octahedra and vice-versa. The Co and Mn ions are arranged alternately either at 2c($\frac{1}{2}$,0,0), or 2d(0,$\frac{1}{2}$,0) in the unit cell. The Co/Mn ordering peak for double perovskite has commonly appeared around 2$\theta$ $\approx$ 20$^{\circ}$\cite{XRD-1}. In the XRD data used in this experiment, this peak intensity is too feeble compared to the most intense peak that arises at 2$\theta$ $\approx$ 33$^{\circ}$. So the background signal diminishes this feature. Therefore, Co and Mn ions are not organized in an ordered arrangement in the unit cell. Any other type of Co/Mn ions arrangement in the unit cell gives rise to partial antisite disorder. It is tough to reveal the actual distribution of Co/Mn atoms on both crystallographic sites by the X-ray diffraction method, as both the cations have very similar scattering cross-sections\cite{XRD-2}. 

\begin{table*}[h]
	\centering
	\caption{The room temperature lattice parameters, bond lengths and bond angles from Rietveld analysis of XRD data.}
	\begin{tabular}{|l c c c c c c c|}
		\hline
		\hline
		\multicolumn{7}{c}{\bf {\chb {Lattice parameters}}}\\
		\hline
		a (\r{A})& b (\r{A})& c (\r{A})& $\alpha$($^{\circ}$)&$\beta$($^{\circ}$)&$\gamma$($^{\circ}$)&Volume (\r{A}$^3$)&$\chi^2$\\
		\hline
		5.2672(3)& 5.5929(4)& 7.5075(5)&90&89.889(10)&90&221.16(3)& 1.08 \\
		
		\hline
		\multicolumn{8}{c}{{\bf{\chb {Bond length and Bond angles}}}}\\
		\hline
		Y/Gd-O (\r{A}) & Mn-O1 (\r{A}) & Mn-O2 (\r{A}) &Mn-O3(\r{A})  &Co-O1 (\r{A}) & Co-O2 (\r{A}) & Co-O3(\r{A})&\\
		\hline
		2.4844& 1.9915& 1.8207& 2.0200& 1.9056 &2.1327 & 1.9700 &\\
		\hline
		\hline
		Co-O1-Mn & & &Co-O2-Mn& & &Co-O3-Mn& \\ 
		\hline
		148.665&&& 152.563 &&& 148.0&\\
		\hline
		\hline
		
	\end{tabular}
\end{table*}

The estimated lattice parameters from Rietveld refinement are tabulated in Table 1. This is seen that the Co-O and Mn-O bond lengths are not similar, suggesting that the CoO$_6$ octahedra are larger in size in comparison to MnO$_6$ octahedra. Again the ionic radii of A site rare-earth ions are different. As a result, the CoO$_6$ and MnO$_6$ octahedra are rotated to balance the misfit of ionic radii. So the bond angle value has strongly deviated from 180$^{\circ}$, a conventional value of ideal double perovskite\cite{XRD-3}. This can be understood by the tilting angle $\phi$ between the CoO$_6$ and MnO$_6$ octahedra, which can be evaluated from $\phi$=(180$^{\circ}$-$\theta$/2)\cite{XRD-4}. $\theta$ is the average bond angle between Co and Mn, and the value of $\phi$ comes out to be 17.6$^{\circ}$. These dissimilarities in bond angle, bond length, and tilting angle $\phi$ indicate a significant distortion present in the structure due to Gd doping. 


\subsection{X-ray photoelectron spectroscopy (XPS)}

In the DP systems, B site cations are often present in mixed-valence states due to several reasons, such as different synthesis conditions, charge compensation, and different type of structures\cite{XPS_0}. The physical and magnetic properties of DP systems are strongly influenced by the oxidation states of the B site cation. Therefore XPS was carried out to find out the oxidation states of B site cations of the YGdCMO sample.
\begin{figure*}[htp]
	\centering
	\includegraphics[scale=.25]{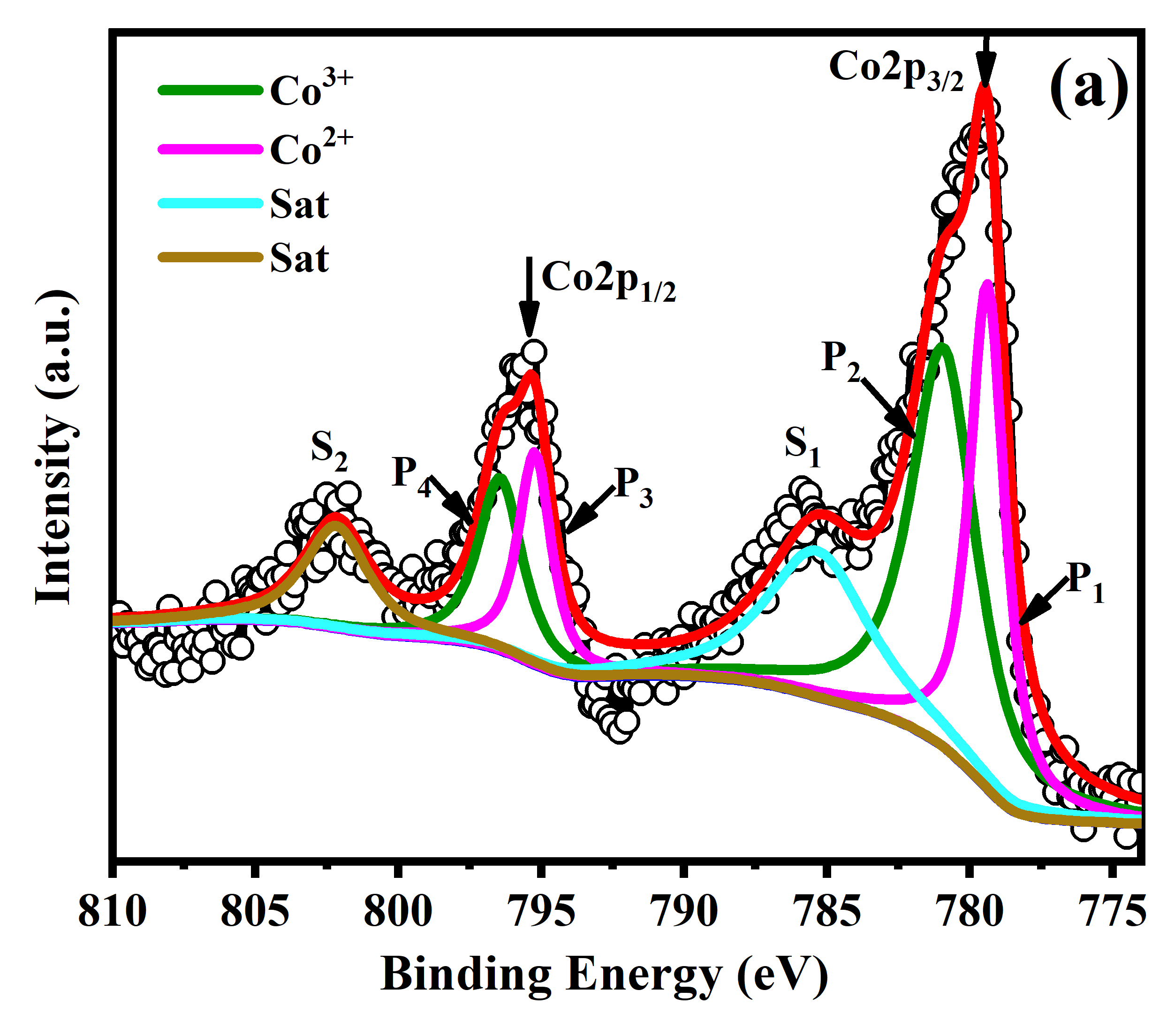}  
	\includegraphics[scale=.25]{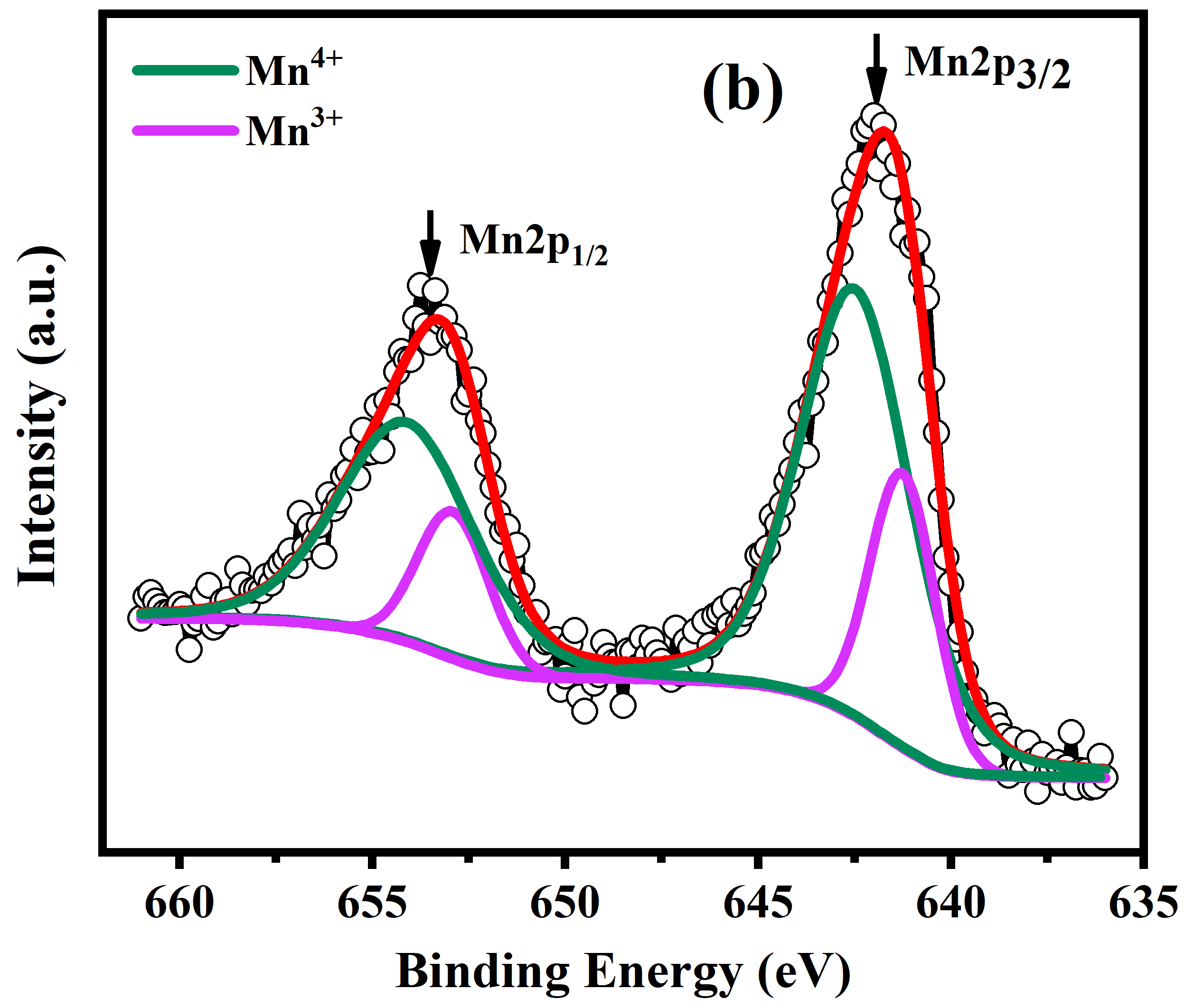}
	\includegraphics[scale=.25]{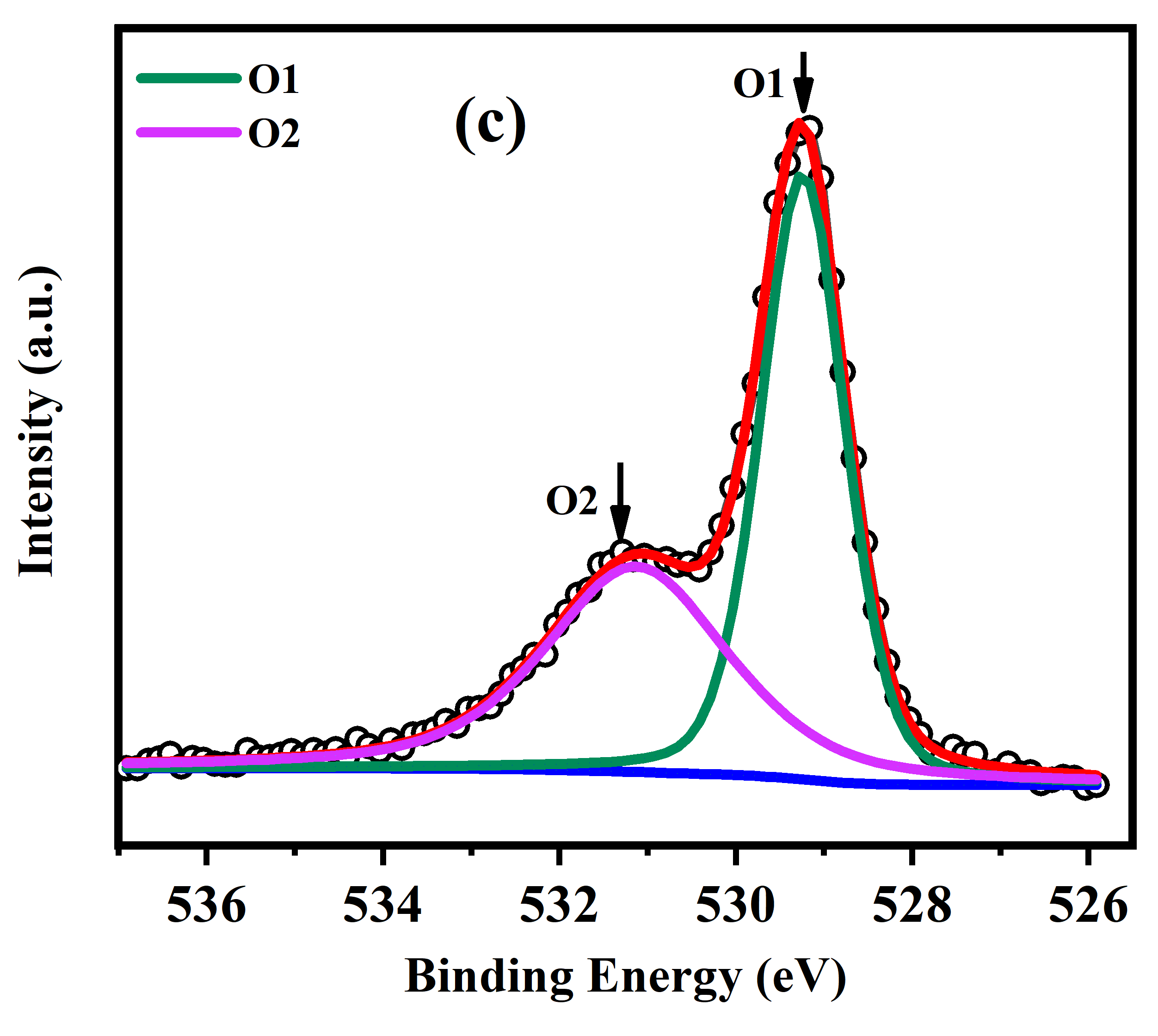}
	\caption{XPS spectra of (a) Co 2$p$, (b) Mn 2$p$ and (c) O 1$s$ powder YGdCMO. Data points are represented by black circles with overall fitted curve shown by the red line. De-convoluted spectral peaks and Shirley background are represented by solid lines of different colors.}
\end{figure*}
\begin{table*}[h]
	\centering
	\caption{Binding energy and the percentage of area obtained by deconvolution of XPS spectra for different oxidation state for Co and Mn ions.}
	\begin{tabular}{l c c}
		\hline
		\hline
		Oxidation  &	Binding  & Area (\%) \\
		state&		Energy (eV)&          \\
		\hline
		\hline
		Co$^{2+}$2p$_{3/2}$ (P$_1$) &	779.38 &39.9 \\
		Co$^{2+}$2p$_{1/2}$ (P$_3$) &   795.22 &\\
		Co$^{3+}$2p$_{3/2}$ (P$_2$) &   780.98 &60.11 \\
		Co$^{3+}$2p$_{1/2}$ (P$_4$) &   796.20 & \\
		Mn$^{4+}$2p$_{3/2}$ (P$_1$) &   642.47 &69.33\\
		Mn$^{4+}$2p$_{1/2}$ (P$_3$) &   654.90 &\\
		Mn$^{3+}$2p$_{3/2}$ (P$_2$) &   641.24 &30.67\\
		Mn$^{3+}$2p$_{1/2}$ (P$_4$) &   652.09 &\\
		\hline
		\hline
		
	\end{tabular}
\end{table*}

 Co-2p core-level spectra consists of two major peaks along with two satellite peaks S$_1$ and S$_2$ illustrated in Figure 2(a). The two major peaks, namely, Co-2P$_{3/2}$ and Co-2P$_{1/2}$ due to spin-orbit splitting are centered at 779.90 eV and 795.15 eV, respectively with a separation of about 15.25 eV\cite{XPS-1}. These peaks are further deconvoluted into four peaks P$_1$, P$_2$, P$_3$, and P$_4$, respectively. It signifies the existence of Co$^{2+}$ in addition to Co$^{3+}$ ion. The obtained binding energy values and area under the peaks are presented in Table 2 and agree well with the literature value\cite{XPS-11, XPS-12}. The satellite peaks S$_1$ and S$_2$ are at the high binding energy side (785.32 eV and 816.12 eV)\cite{XPS-2}. This is a clear indication of the existence of mixed oxidation states of Co ions. The energy difference ($\Delta$E) between two-level, due to spin-orbit coupling was found to be different for divalent ion and trivalent ion with $\Delta$E $\sim$ 15.5 - 16 eV for Co$^{2+}$ and 15.0 - 15.2 eV for Co$^{3+}$ ion, respectively\cite{XPS-4, XPS-5}. In the present experiment, it is found to be 15.8 eV and 15.2 eV for Co$^{2+}$ and Co$^{3+}$, respectively. It confirms that the Co$^{2+}$ and Co$^{3+}$ ions are present in our sample. The corresponding ratio of Co$^{2+}$ and Co$^{3+}$ ions is $\sim$ 4:6.\\
 
 Mn-2p spectra associated with two major peaks are located at 641.80 eV and 653.37 eV, respectively. The energy difference between spin-orbit splitting 2p$_{3/2}$ and 2p$_{1/2}$ peaks is 11.57 eV\cite{XPS-6, XPS-7}. These peaks are deconvoluted into four consecutive peaks, namely, P$_1$, P$_2$, P$_3$, and P$_4$, indicating the presence of Mn$^{4+}$ ion along with Mn$^{3+}$ ion shown in Figure 2(b). The binding energy value and area under the fitted peaks are summarized in Table 2. The observed binding energy value for Mn$^{4+}$ and Mn$^{3+}$ are in good agreement with the reported value\cite{XPS-8, XPS-9}. The full width at half maxima (FWHM) plays an important role in the determination of the monovalent or multivalent oxidation states of the B site cation. In our sample, the FWHM of Mn-2p$_{3/2}$ peak is found to be 5.1 eV, which is greater than that of the divalent and trivalent oxidation state of Mn in MnO$_2$ (3.2 eV), Mn$_2$O$_3$ (3.0 eV)\cite{XPS-10}. This large value of FWHM signifies coexistence of multivalent Mn ions in our sample. The ratio of the Mn$^{4+}$ and Mn$^{3+}$ ions are equal to the ratio of the peak area and the obtained value is $\sim$ 7:3.\\
 
 Figure 2(c) represents the XPS spectra of O 1s core level. The deconvoluted peaks O1 and O2 are centered at 529.24 eV and 531.13 eV, respectively. The O1 peak comprises lattice oxygen bonding with metal (Co, Mn), whereas the O2 peak is associated with absorbed oxygen on the surface\cite{XPS-13, XPS-14}.

\subsection{Magnetic measurements}

\subsubsection{Temperature dependent DC magnetization}
\begin{figure*}[htp]
\centering
\includegraphics[scale=.75]{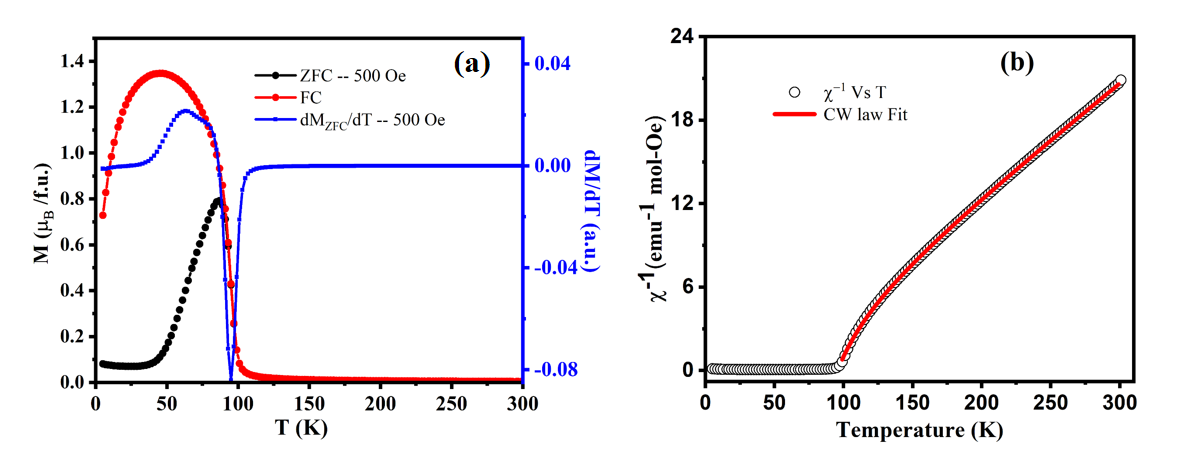}  
\caption{(a) M(T) curve of YGdCMO measured at 500 Oe dc magnetic field H$_{dc}$ under ZFC and FC and its temperature derivative dM$_{ZFC}$/dT. (b) Modified CW law fit to the experimental data of $\chi^{-1}$.}
\end{figure*}
   
It was observed previously that the distortion and site disorder present in the DP system have a significant impact on its magnetic properties\cite{int_12}. To elucidate the evolution of magnetic ground states with site disorder and distortion, temperature-dependent FC and ZFC magnetization M(T) measurements have been carried out.

Figure 3(a) shows the thermal variation (T) of ZFC and FC magnetization (M(T)) of Gd-substituted YCMO under the application of 500 Oe dc magnetic field. Three magnetic transitions have been observed in the M-T curve, and corresponding ordering temperatures are determined by the first-order derivative of [dM$_{ZFC}$/dT] shown in Figure 3(a). 
i) A sharp change in magnetization of FC and ZFC curves have been observed at Curie temperture \textit{T}$_C$=95.5 K. This paramagnetic to ferromagnetic ordering is resulting from spin- spin superexchange interaction\cite{mag_3} between Co$^{2+}$ and Mn$^{4+}$. The virtual electron transfer between half-filled orbital (e$_g^{2}\uparrow$) of Co$^{2+}$(3d$^7$: t$_{2g}^{3}\uparrow$ t$_{2g}^{2}\downarrow$ e$_{g}^{2}\uparrow$)to empty orbital (e$_g^0$) of Mn$^{4+}$(3d$^3$: t$_{2g}^{3}\uparrow$ e$_{g}^{0}$) gives strong ferromagnetic coupling in this sample\cite{mag_1}.

ii) A substantially significant bifurcation between FC and ZFC curves has been observed below \textit{T}$_C$. This feature is common for cobaltite and dilute doped manganite systems, indicating strong anisotropy and the existence of mixed phases at a lower temperature\cite{mag_4}. Below \textit{T}$_C$, the M$_{FC}$ magnetization value reaches a maximum at 50 K, then starts decreasing around \textit{T}$_N$= 47 K followed by a broad peak, which indicates non-zero polarisation of magnetic ions with AFM-like behavior. This AFM transition can be associated with strong 3d-4f exchange interactions between FM sublattice of Co$^{2+}$/Mn$^{4+}$(3d$^7$/3d$^3$) and Gd$^{3+}$(4f$^7$) ion\cite{mag_2} and due to ASD present in our sample.

iii) Below \textit{T}$\leq$20 K, an up-turn anomaly in M$_{ZFC}$ curve is visible, whereas M$_{FC}$ curve shows a sharp downfall. This feature is more pronounced with an increase in dc magnetic field shown in Figure. 4(a). This can be attributed to the gradual development of Gd spins ordering along the \textit{ab} plane below this temperature which is initially in \textit{ac} plane. This result is coherent with our first-principles calculation. The short-range exchange interaction of Gd$^{3+}$-O-Gd$^{3+}$ gives rise to an AFM ordering of Gd spin. Similar feature was observed in Ho$_2$NiMnO$_6$\cite{AC_1}.

Furthermore, in comparison with the parent compound YCMO\cite{mag_5}, the substantial increase of magnetization value of Gd-substituted YCMO is observed below \textit{T}$_C$ due to magnetic Gd$^{3+}$ ion present in the sample. The Curie temperature \textit{T}$_C$ is one of the important parameter, which holds its potential application in real devices. The higher value of  \textit{T}$_C$ makes it more applicable in the industry. The larger size of Gd$^{3+}$ ion would help to straighten the average bond angle of Co-O-Mn and it contributes to large electron density at the Fermi level. This will increase the electron hopping ability, consequently superexchange interaction between Co$^{2+}$ and Mn$^{4+}$ ions via intervening oxygen. Hence the \textit{T}$_C$ of the Gd-substituted compound increases to 95.5 K whereas it is 75 K for parent compound\cite{mag_5}.  \\
\\
 The temperature-dependent inverse of susceptibility has been displayed in Fig. 3(b) to delineate the nature of the magnetization in high-temperature regions (i.e. in PM state). The non-linear nature of the curve above \textit{T}$_C$ implies that the conventional Curie-Weiss law deviates in the PM region. The bulging nature of the curve signifies that the total magnetic contribution to the susceptibility is the sum of the contribution of transition metal sublattices and rare-earth sublattice site\cite{AC_1}. Here it is assumed that the rare earth and transition metal paramagnetism are noninteracting in nature. It is possible due to localize nature of heavy rare earth ion Gd. Therefore the modified Curie-Weiss law with the following expression describes the magnetic behavior in YGdCMO. 
\begin{equation}
    \chi=\frac{C_{Co-Mn}}{T-\theta_{TM}}+\frac{C_{Gd}}{T-\theta_{Gd}}
\end{equation}
where C$_{Co-Mn}$ and C$_{Gd}$ are the Curie constants of Co-O-Mn sublattice and Gd sublattice, respectively, and $\theta_{TM}$ and $\theta_{Gd}$ are the Curie-Weiss temperature for transition metal ions and rare earth (Gd) ion. The Curie constant is related to $\mu_{eff}$ by the formula
\begin{equation}
    \mu_{eff} = \sqrt{\frac{3k_{B}C}{N\mu^2}} = 2.827\sqrt{C}
\end{equation}
where k$_B$ is the Boltzmann constant, $\mu_B$ is the Bohr magnetron, and N is the number of magnetic atoms per unit volume. The $\chi^{-1}$ vs. T plot is shown in Figure 3(b), and the curve is well fitted to Eq. (1), indicating by the solid red line. The obtained values of fitting parameters $\theta_{TM}$ is 95.68 K which is positive and close to \textit{T}$_C$ and C$_{Co-Mn}$ 3.93 emu K mol$^{-1}$Oe$^{-1}$. Therefore the value of effective paramagnetic moment $(\mu_{eff})_{Co-Mn}$ is estimated from Eq.(2), and it is found to be 5.60 $\mu_B$/f.u. The theretical effective paramagnetic moment described as $\mu_{theo}=\sqrt{\mu_{Co}^2+\mu_{Mn}^2} $ where $\mu_{Co}= g\sqrt{S(S+1)}$ and $\mu_{Mn}= g\sqrt{S(S+1)}$ with $g\approx$2.0. The effective moment value for different spin configurations are given follows: 5.47 $\mu_B$ for [Co$_{HS}^{2+}$=3/2 and Mn$_{HS}^{4+}$=3/2 (HS=High spin)]; 5.65 $\mu_B$ for [Co$_{IM}^{3+}$ (Intermediate spin)=1 and Mn$_{HS}^{3+}$=2]; 6.92 $\mu_B$ for [Co$_{HS}^{3+}$=2 and Mn$_{HS}^{3+}$=2]. Therefore, the observed experimental moment value indicates that the existence spin interaction mediated through mixed valence state.  
For Gd sublattice, the extracted values of $\theta_{Gd}$ and C$_{Gd}$ are 20 K and 9.06 emu K mol$^{-1}$ Oe$^{-1}$, respectively. The positive value of $\theta_{Gd}$ indicates the FM ordering of Gd sublattice at a lower temperature. The calculated value of $(\mu_{eff})_{Gd}$ from Eq.(2) comes out to be 8.51 $\mu_B$/f.u.. This value is higher than the effective paramagnetic moment for free Gd$^{3+}$ ion (7.0 $\mu_B$/f.u.). The model of modified Curie-Weiss law is based on the non-interacting Co/Mn and Gd sublattice. So, this higher value might be due to the interaction between Co/Mn and Gd ions at lower temperature present in our sample.

The effective paramagnetic moment for a perfectly ordered, well-known system (LCMO) is 6 $\mu_B$\cite{int_9}, as reported earlier. In comparison with ordered LCMO the effective moment of YCMO is a little bit lower. The 3d-4f exchange interaction between Co/Mn sublattice and Gd spins plays a crucial role in producing AFM interaction in the system. Moreover, the site disorder present in the sample leads to the formation of mixed valences of Co and Mn ions. As a result, different short-range AFM interactions such that Co$^{3+}$-O-Co$^{3+}$, Mn$^{3+}$-O-Mn$^{3+}$ come into play along with FM interaction. So the overall magnetic moment of the system decreases. However, in a perfectly ordered system, only Co$^{2+}$-O-Mn$^{4+}$ superexchange interaction exists, which is responsible for FM interaction and gives the higher value of the moment. Approximately we have calculated the percentage of antisite disorder present in our sample by using Eq. (3)\cite{mag_6}, where M$_{Co}$ and M$_{Mn}$ are the spin-only moment, and $x$ is the amount of hole doping.
\begin{equation}
	M_S=(1-2ASD)[M_{Co}+M_{Mn}]+x(1-2ASD)
\end{equation}
The 1st term in the equation corresponds to the contribution of ASD to the M$_S$ and the 2nd term is indicated the reduction of magnetization due to the doping element. The value comes out to be 38\%, which is quite considerable.\\
\\

\begin{figure*}[htp]
	\centering
	\includegraphics[scale=.75]{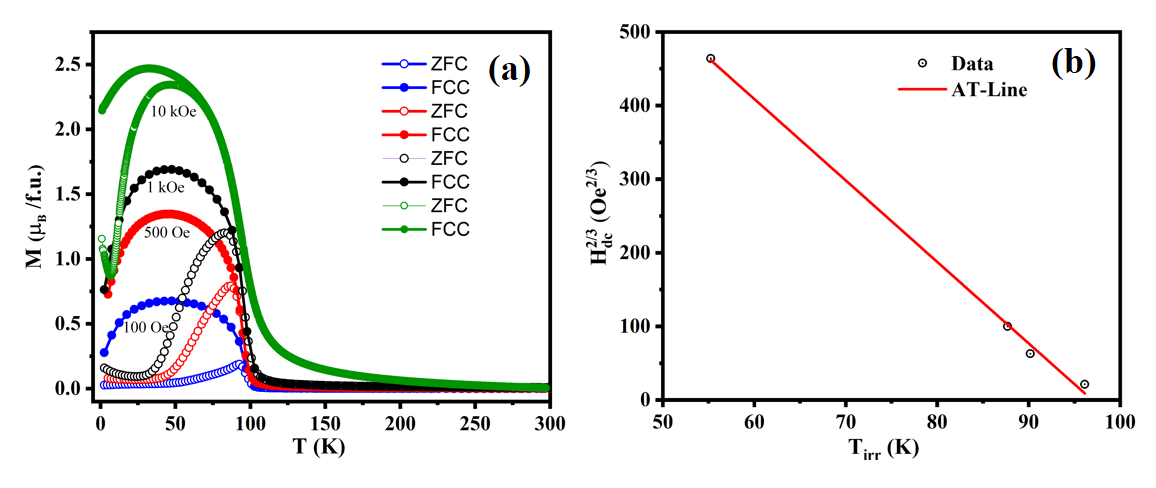}
	\caption{(a) Temperature dependent magnetization under the application of various dc fields. (b) The H$^{2/3}$ Vs T plot along with the fitting of Almeida Thouless line.}
\end{figure*}
To further study, temperature-dependent magnetizations have been measured under the application of 100 Oe, 1 kOe and 10 kOe magnetic fields. All the ZFC and FC curves follow the same pattern. It is seen that there is still a bifurcation between the ZFC and FC curve under the application of even 10 kOe field. It means that this field is not sufficient to align all the spin along the field direction. This is a necessary condition (not sufficient) for spin-glasses to be present in the system. There is irreversibility (at which temperature the deviation between ZFC and FC curve takes place) observed below 100 K. Previously this type of feature was ascribed to spin-glass or SPM-like behavior. But the AC susceptibility measurement rules out the possibility of having spin-glass in the system. Figure 4(a) depicts the shifting of T$_{irr}$ towards the lower temperature with an application of a high field, indicating that the frozen state is relaxed with the help of an external magnetic field. The variation of T$_{irr}$ with the external magnetic field can be mapped onto the De-Almedia thoughtless (AT)-line\cite{mag_7,mag_8}, which is given below
\begin{equation}
	H_{dc}(T_{irr})=\Delta J[1-\frac{T_{irr}(H_{dc})}{T_{SG}}]^{3/2}
\end{equation}
Here $\Delta J$ is associated with exchange interaction, T$_{SG}$ is the SG freezing temperature at zero applied field. The experimental curve is nicely fitted by Eq. 4 shown in Figure 4(b), and the slope of this straight line gives the spin freezing temperature T$_{SG}$=96.94 K. This indicates the volume spin glass-like behavior\cite{mag_9,mag_10}.


\subsubsection{AC susceptibility}

\begin{figure*}[htp]
	\centering
	\includegraphics[scale=.96]{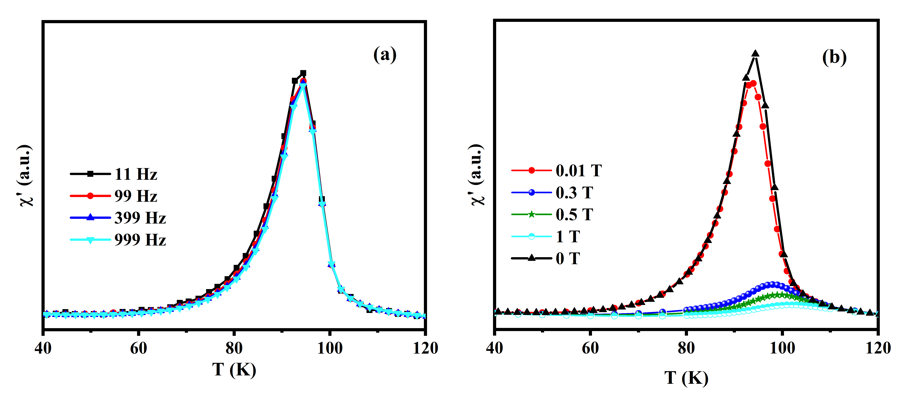}  
	\caption{(a) Temperature dependent AC susceptibility for different frequencies and (b) different dc fields superimposed.}
\end{figure*}
The temperature dependent AC susceptibility measurements, with an AC field of 1 Oe and different frequencies, have been carried out to probe the presence of glassy magnetic phase in the YGdCMO sample. The sharp peak corresponding to ferromagnetic transition is observed nearly at 94.82 K, shown in Figure 5(a), whereas no anomaly is noticed corresponding to AFM transition. All the peaks are being frequency independent, which rules out the possibility of having spin-glass phase in our sample. Moreover, the memory effect experiment was performed (following the protocol\cite{AC_2}) below the transition temperature (plot not shown here). No memory effect is observed near the waiting temperature, which is in disfavor of the presence of SG/cluster glass. The AC susceptibility with the variation of DC field is shown in Figure 5(b). The curve depicts that the susceptibility maxima shifts towards higher temperature with increase in DC magnetic field. This is ascribed to the presence of critical fluctuations along with a continuous transition to a ferromagnetic state.

Earlier it was reported that the $\chi^{\prime}$ remains constant down to low temperature for a purely FM material. However, the $\chi^{\prime}$ decreases below the transition temperature for many inhomogeneous magnetic oxide and amorphous alloys, having a finite size of FM and AFM clusters. This is associated to magnetic anisotropy\cite{AC_3}. The inhomogeneous FM clusters in many FM systems, change their shape and size continuously below \textit{T}$_C$ lead to magnetic anisotropy. The sharpness of the susceptibility peak and decrease in amplitude of susceptibility strongly depends upon the disorder present in the sample, spin-spin correlation, and anisotropy energy. This anisotropy energy blocks the spin and do not allow them to respond in a weak field below \textit{T}$_C$.
\begin{equation}
	\chi'(T)\propto M_S'^2(T)/K(T)
\end{equation} 
From Eq. 5 it can be explained, where M$^{\prime}_S$ is the spontaneous magnetization and K(T) is the anisotropy energy at a particular temperature\cite{AC_4}. The FM spin-spin correlation increases with the decrease in temperature. Therefore, the spontaneous magnetization as well as anisotropy energy increase below \textit{T}$_C$ which results decrease in susceptibility.

Therefore, the sharp peak in the susceptibility curve arises due to ASD present in our sample as externally no disorder is added to the system. But ASD is very common in DP system. So, ASD can not alone contribute to this sharp peak. In our sample different interactions Co-O-Mn, Co-O-Co, Mn-O-Mn are arise due to FM and AFM interaction, leading to magnetic frustration. Thus the sharp peak could be result for both ASD and magnetic frustration.  
\subsubsection{Magnetization study using first principles DFT }

\begin{figure*}[htp]
	\centering
	\includegraphics[scale=1.0]{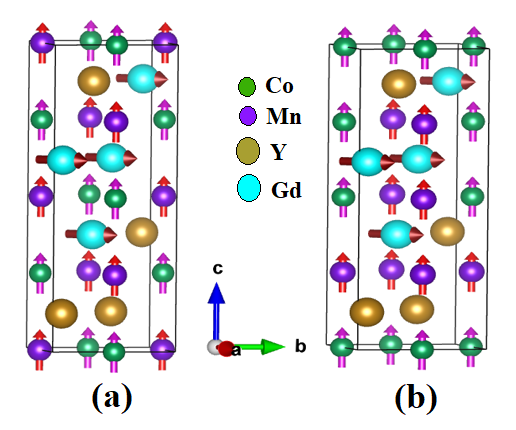}
	\caption{  Schematic spin structure of room temperature (a) ordered YGdCMO (b) disorder YGdCMO (right pannel). }
\end{figure*}
In order to support experimental findings, we carried out first-principles DFT calculations. We have considered supercells of size 1×1×2 containing 40 atoms. The total-energy calculations were performed to obtain the ground-state of the system using the relaxed structural parameters (starting from experimentally obtained crystal structure) for various magnetic configurations within the generalized gradient GGA+U and GGA+U+SO (which includes spin-orbit interaction) approximations. The value of U$_{eff}$ = U-J (where U is the Coulomb correlation and J is Hund’s exchange) used for Gd, Mn, and Co are 6.1 eV, 3.1 eV, and 4.1 eV, respectively. Figure 6(a) shows an ordered arrangement of YGdCMO where Co and Mn atoms are in alternate positions. Figure 6(b) Co and Mn atoms are randomly distributed, known as a disordered structure. 

For the collinear calculations with spin moments of Gd, Mn, and Co pointing along the \textit{c} direction, we have considered six different spin configurations to determine the magnetic ground state of the system: a) AFM configuration Gd (↑ ↑ ↑ ↑) Mn/Co (↑ ↓ ↑ ↓), b) FM configuration Gd, Mn/Co all pointing in same direction c) Ferrimagnetic configuration – Gd (↑ ↑ ↑ ↑), Mn/Co (↓ ↓ ↓ ↓). In the other three spin configurations, we have kept Mn/Co spin in the up direction, while for Gd, we have taken G-type AFM, C-type AFM, and A-type AFM spin configurations. The relative energy values are given in table 3. We observe that the ground state magnetic order is ferromagnetic. The corresponding moment values obtained from DFT calculations for Gd, Mn, and Co are 7.05 $\mu_B$, 3.1 $\mu_{B}$ and 2.6 $\mu_B$, respectively. Therefore, in an ordered structure Co and Mn ions are close to the nominal valence of +2 and +4, respectively, which corroborate the results of XPS analysis for ordered structure.
\begin{table*}[h]
	\centering
	\caption{Relative energy per formula unit (in eV) of various spin configurations within GGA+U+SO approximation.}
	\begin{tabular}{l c}
		\hline
		\hline
		Spin configurations& Energy values (eV)\\
		\hline
		\hline
		AFM  &	0.095\\
		FM	&0.0\\
		Ferrimagnetic &	0.10 \\
		G-type (Gd) &	0.03 \\
		C-type (Gd)	& 0.055 \\
		A-type (Gd)	 & 1.18 \\
		\hline
		\hline
		
	\end{tabular}
\end{table*}

We further computed total energies assuming noncollinear spin configurations within the GGA +U + SO approximation to find out the preferred direction of the Gd spin moments. We have chosen four different directions for Gd moments, while keeping the Co and Mn spin moments fixed along \textit{c} direction in FM spin configuration: a) Gd spins along \textit{a} direction b) Gd spins along \textit{b} direction c) Gd spins along \textit{c} direction, and d) Gd spins moments confined in \textit{ab}-plane pointing along [110] direction. The relative energies are given in table 4. It is found that Gd moments are lying in \textit{ab} plane.
\begin{table*}[h]
	\centering
	\caption{Relative energy (in eV) of various spin configurations within GGA+U+SO approximation.}
	\begin{tabular}{l c }
		\hline
		\hline
		Spin configurations of Gd &	Energy values (eV) \\
		\hline
		\hline
		(1, 0, 0) &	0.015 \\
		(0, 1, 0) &	0.04 \\
		(0, 0, 1) & 0.035 \\
		(1, 1, 0) &	0.0  \\
		
		\hline
		\hline
		
	\end{tabular}
\end{table*}

We have also performed the non-collinear calculation with the anti-site disordered structure of YGdCMO to investigate the ionic state of Mn/Co atoms. The corresponding moment values obtained from DFT calculations for Mn and Co are 3.8 $\mu_{B}$ and 1.9 $\mu_{B}$, respectively. It implies that the Mn, and
	Co ions are in +3 oxidation state which is in good agreement with the XPS result. We observed that the magnetic ground state of the system comprises of Co and
	Mn moments being along \textit{c} direction, while Gd moments residing in \textit{ab} plane, which is consistent with our experimental findings of DC magnetization. Also, it is seen that the Co and Mn ions change their oxidation state due to the effect of disordered. Both Co and Mn are in +3 state for disordered structure, whereas Co and Mn are in +2 and +4 oxidation state, respectively for the ordered structure. The moment of Co ion decreases which implies short-range interaction with its neighbors in disordered state. In that case Co-3d and O-2p hybridization increases. For Mn ion, the moment value increases due to decrease in hybridization between Mn-3d and O-2p orbitals, indicating long-range ordering. Therefore, there is a competition between long-range and short-range interactions which is responsible for observed magnetic behavior in our system.

\subsubsection{Isothermal magnetization}
To elucidate the magnetic properties of the Gd-substituted YCMO with the variation of the external magnetic field, isothermal magnetization M(H) measurements have been performed up to a field $\pm$7T at 5K, 10 K, 50 K and 100 K, respectively in ZFC mode. Figure. 7(a) shows the M(H) curve measured at 5 K. Inset of this Figure represents the virgin curve taken at 5 K from 0 to 7T of ZFC M(H). The Figure. 7(b) represents the M(H) curves measured at 10 K, 50 K, and 100 K, respectively. The M(H) loop at 5 K has a sizeable coercive field H$_{cr}$ 9.45 kOe and magnetic coercivity 0.73 $\mu_B$/f.u. shows FM-like behavior. At 100 K, the M(H) loop is a perfect straight line with zero coercive field and magnetic coercivity indicating that the sample behaves as a perfect paramagnet and is free from any magnetic impurity. Figure. 7(b) depicts that magnetization value increases with a decrease in temperature. The localized Gd-4f moments would help to increase the magnetization value at a lower temperature. It is clearly seen from the hysteresis loops, measured at 5K, 10K, 50K and 100K that the magnetization does not saturate even at H=7T magnetic field. This feature is obvious for rare-earth sublattice present in the compound. This incomplete saturation of magnetization corresponds to the $3d$-$4f$ exchange interaction and ASD present in the sample. So the incomplete saturation of magnetization at high field and low field hysteresis loops (in the temperature range 5 K-50 K) indicate the existence of FM/AFM phases.
\begin{figure*}[htp]
	\centering
	\includegraphics[scale=.7]{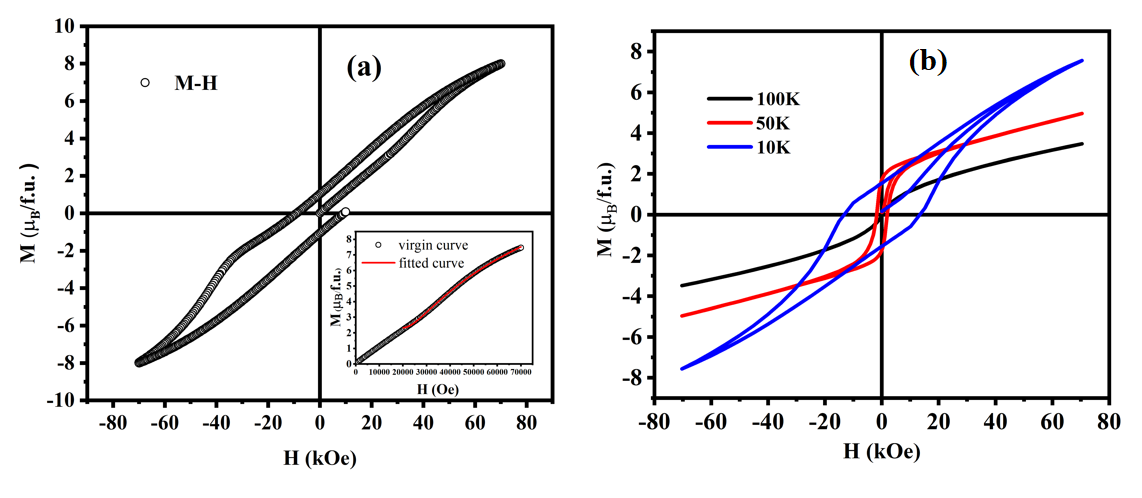}  
	\caption{(a) ZFC M(H) isotherms in the range ±70 kOe at 5 K. Inset shows the virgin curve isotherm taken at 5 K. (b) Isothermal hysteresis loops are measured at different temperatures 10 K, 50 K and 100 K.}
\end{figure*}

So from the virgin M(H) curve shown in the inset of Figure. 7(a) taken at 5K, we have estimated saturation magnetization value by fitting the high magnetic field part (H$\geq$20 kOe) using the Eq.6\cite{mag_11} where c is the high field differential susceptibility, a comprised with the structural defect and nonmagnetic inclusion of local magnetic moments and b is associated to magnetocrystalline anisotropy of the system. 
\begin{equation}
	M(H)= M_S[1-\frac{a}{H}-\frac{b}{H^2}]+cH
\end{equation}
From the fitting, the obtained value of M$_S$ is 9.14 $\mu_B$/f.u., which is deviated from the sum of the theoretically estimated value of fully polarised Co$^{2+}$-O-Mn$^{4+}$ (6.0 $\mu_B$/f.u.) and Gd spin (7.0 $\mu_B$/f.u.) system. So it is seen that the obtained saturation magnetization value is much lower than the theoretically obtained value. This decrease in the moment is a clear impact of ASD present in our sample\cite{mag_14} because ASD creates ABPs, which are also responsible for reducing the saturation magnetization value\cite{XRD-4, AC_1,mag_13}. The presence of a strong magnetic pinning force with disorder restricts the moments to align along the field direction\cite{mag_12}. Thus the magnetization value decreases. This disordered structure generates several types of AFM interaction such as Co$^{2+}$-O-Co$^{2+}$, Co$^{3+}$-O-Co$^{3+}$, Co$^{2+}$-O-Co$^{3+}$, Mn$^{3+}$-O-Mn$^{3+}$, Mn$^{4+}$-O-Mn$^{4+}$ and Mn$^{3+}$-O-Mn$^{4+}$ along with the Co$^{2+}$-O-Mn$^{4+}$ FM interaction. This AFM interaction helps to reduce the overall magnetization value. However, there is a direct dependence between the ASD and reduction of the magnetic moment\cite{mag_14}. Fig. 7(b) depicts that at low temperatures, our sample exhibits large coercivity. This is ascribed to the pinning of the FM domain wall due to ASD defects and magnetocrystalline anisotropy of the sublattices\cite{mag_13}. Again at high temperatures the thermal energy suppresses the magnetocrystalline anisotropy energy, and the domain wall starts moving, as a result, the coercivity decreases\cite{mag_13}. As the YGdCMO compound contains mixed magnetic phases due to ASD, metamagnetic-like transitions are expected which reflects in the nature of the hysteresis curves. Previously this type of phenomenon is observed in RCo$_{1-x}$Mn$_x$O$_3$ (R=Eu, Y, Nd)\cite{mag_15}. Therefore it is seen that ASD has a more significant influence on saturation magnetization and also some other magnetic properties.

\subsection{Exchange bias effect}

The competition between different magnetic interactions and uncompensated moment at pinning boundary of FM/AFM phases in oxide sample plays a significant role in engendering the exchange bias (EB) phenomena\cite{EX_1}. The coexistence of mixed valences of Co and Mn ions, $3d-4f$ exchange interaction, and site disorder in the YGdCMO compound are responsible for producing FM and AFM exchange interactions at low temperature, resulting a significant amount of frustration, and anisotropy at the FM/AFM interface in the system. These are the adequate conditions to observe EB in oxide samples. So we could expect similar kinds of phenomena in the YGdCMO compound. To understand this feature properly ZFC and FC hysteresis loops are measured at 5 K, after cooling the system from PM region to measure temperature with the different cooling fields.
\begin{figure*}[htp]
\centering
\includegraphics[scale=.7]{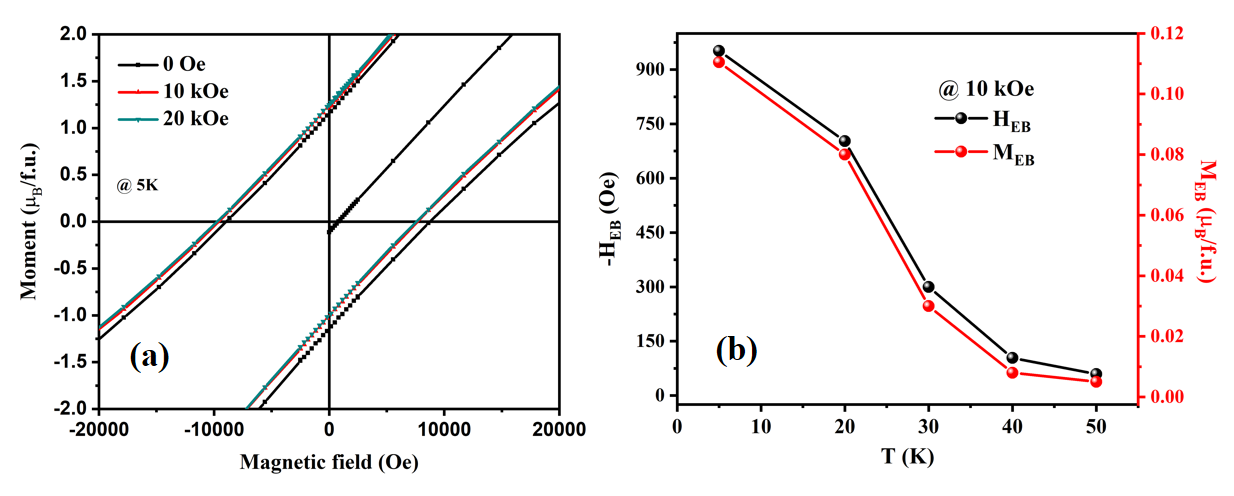}\\
\includegraphics[scale=.8]{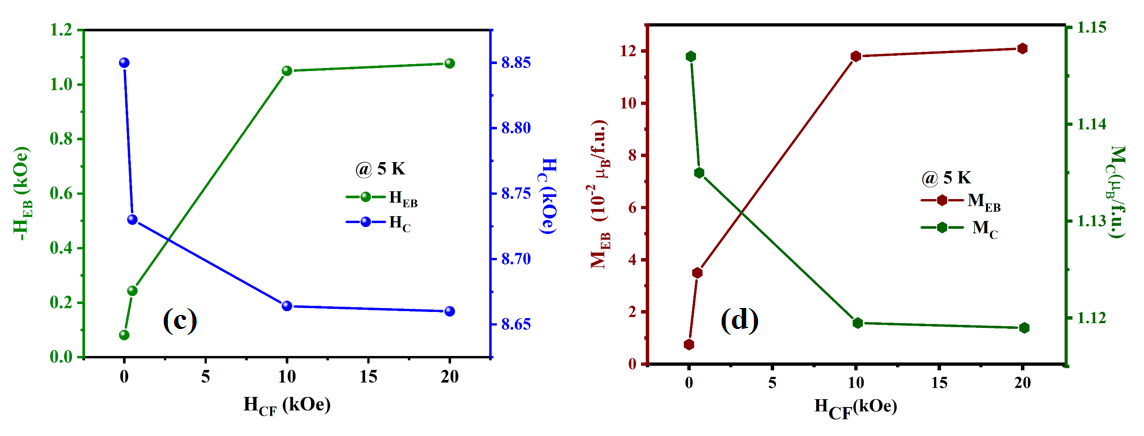}
\caption{(a) M(H) isotherms measured at different cooling fields at 5 K. (b) Variation of \textit{H}$_{EB}$ and \textit{H}$_{C}$ with temperature. (c) \textit{H}$_{EB}$ and \textit{H}$_{C}$ dependence with cooling field at 5 K. (d) Variation of \textit{M}$_{EB}$ and \textit{M}$_C$ with cooling field at 5 K.}
\end{figure*}
The minimum anisotropy field required to flip the spin should be known to get the conventional exchange bias effect is approximately estimated by using the fitting parameter b in Eq. 6. The value becomes 500 Oe\cite{EX_2}. Figure. 8(a) depicts the zoomed view of three hysteresis loops, measured at 5 K after cooling the sample with three different external magnetic fields (0 kOe, 10 kOe and 20 kOe). A considerable amount of shift in hysteresis loops is seen along the negative field and positive magnetization direction. This is the manifestation of conventional exchange bias\cite{EX_1}. Zero field cooled loop shift is not noticeable in this compound which excludes the possibility of occurrence of spontaneous EB effect. In terms of loop shift along the field direction as well as the magnetization direction, the EB field is defined by the average of the zero magnetization intercept [\textit{H}$_{EB}$=$\frac{H_{C1}+H_{C2}}{2}$] and coercive field is defined by half of the width of the hysteresis loop at the average of zero field intercepts [H$_C$=$\frac{\mid {H_{C1}}\mid+\mid{H_{C2}}\mid}{2}$], EB magnetization is given by [\textit{M}$_{EB}$=$\frac{M_{r1}+M_{r2}}{2}$] and magnetization coercivity is defined by [\textit{M}$_{C}$=$\frac{\mid{M_{r1}\mid}+\mid{M_{r2}}\mid}{2}$] where H$_{C1}$ and H$_{C2}$ are the left and right coercive field\cite{mag_13}. Similarly, M$_{r1}$ and M$_{r2}$ are the positive and negative remanent magnetization. Fig. 8(c) shows the variation of \textit{H}$_{EB}$ and \textit{H}$_C$ with cooling field. \textit{H}$_{EB}$ increases rapidly with the cooling field and reaches saturation at 20 kOe, whereas \textit{H}$_C$ tends to decrease. The maximum shifts observed in EB and coercivity are 1.07 kOe and 8.85 kOe at 20 kOe and 0 Oe fields, respectively. The dependence of \textit{M}$_{EB}$ and M$_C$ with a cooling field is shown in Figure 8 (d). The \textit{M}$_{EB}$ increases with the cooling field and attains saturation at 20 kOe, while M$_C$ decreases with the cooling field.\\

Thus it is seen that the EB effect generally occurs in a spin-glass type system or a frustrated DP system\cite{EX_1}. In our sample spin-glass phase is ruled out. The long-range FM ordering is getting affected by the AFM phases originating from the site disorder and 3d-4f exchange interaction due to Gd doping in the compound\cite{EX_3}. It would contribute a net unidirectional anisotropy at the FM/AFM interface. Therefore the uncompensated AFM moments with different anisotropies get pinned into the FM matrix, resulting in exchange bias phenomena\cite{EX_4, EX_5}. However, the net value of unidirectional anisotropy is zero in the absence of any applied cooling field, ensuing no loop shift. Now we try to explain the variation of \textit{H}$_{EB}$ with the cooling field with the help of the magnetic ground state. Below $\leq$10 kOe field, a rapid increment of \textit{H}$_{EB}$ is due to large unidirectional anisotropy at the pinning boundary in comparison with the size of the FM domain as well as total FM magnetization. The \textit{H}$_{EB}$ value is found to be saturated above 10 kOe indicating that strong interfacial magnetic unidirectional anisotropy exists at the interface. Generally \textit{H}$_{EB}$ and \textit{M}$_{EB}$ decrease with high cooling field where the EB is driven by SG/FM interface\cite{EX_5}. Also, the decrements of M$_C$ and H$_C$ with the cooling field support our statement.    \\  
\\

To further investigate the origin of EB, the temperature dependence of EB is a vital tool. Hysteresis loops were measured at different temperatures with a fixed cooling field of 10 kOe. The variation of \textit{H}$_{EB}$ and H$_C$ with temperature is presented in Figure 8(b). \textit{H}$_{EB}$ and H$_C$ both are decreased with increase in temperature and the value of \textit{H}$_{EB}$ and H$_C$ almost reaches zero near \textit{T}$_N$\cite{EX_5}. So the EB arises only below \textit{T}$_N$ $\leq$47 K, which indicates that the existence of AFM phases are the reason to observe EB in this compound.\\ 
\\

\begin{figure*}[htp]
\centering
\includegraphics[scale=.9]{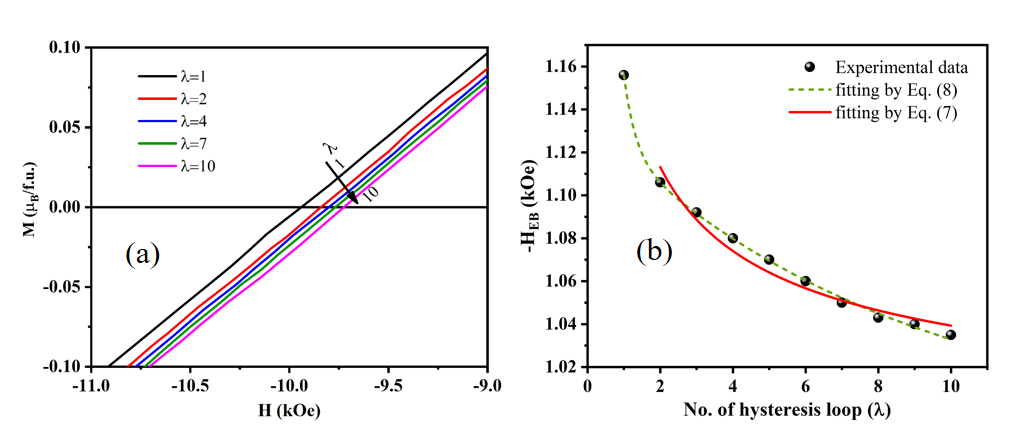}
\caption{(a) Zoomed-in view of the hysteresis loops showing the TE effect of EB. (b) The variation of \textit{H}$_{EB}$ as a function of loop number ($\lambda$) evaluated from hysteresis loops measured at 5 K. The solid and dashed lines represent the best fit for two different models mentioned in the text. }
\end{figure*}

The training effect (TE)\cite{EX_5} is another important feature to unravel the origin of intrinsic EB. The variation of the EB switching field with consecutive hysteresis loops cycled across a particular cooling field is known as the training effect. The EB value decreases with successive field cycling due to rearrangement of spin towards equilibrium\cite{EX_5, EX_6}. The smaller the value of TE is better for application. 
So ten successive cycles have been measured repeatedly at 5 K, after cooling the sample with a 10 kOe field. Figure 9(a). represents the enlarged view of the hysteresis loops in the negative field regime (only a few hysteresis loops are shown for a better view). Figure 9(b) reveals that the EB value decreases with an increase in the loop no. $\lambda$, indicating the presence of TE in the YGdCMO compound. In our sample, the decrements between the 1st and 2nd hysteresis loop are small, indicating that the interfacial spins are more stable against the applied external magnetic field. The dependence of H$_{EB}$ with loop number can be described by following power law\cite{EX_7}.
\begin{equation}
	H_{EB}(\lambda)-H_{EB}(\infty) \propto \frac{1}{\sqrt \lambda}
\end{equation}
where \textit{H}$_{EB}$($\infty$) is the EB field at $\lambda \rightarrow$ $\infty$.
The fitted curve for $\lambda$=2 to $\lambda$=10 is represented by the solid red line shown in Figure 9(b). The approximate value of obtained fitting parameter \textit{H}$_{EB}$($\infty$) is 979 Oe. The power-law explains only the evolution of free energy due to the deviation of interfacial AFM magnetization\cite{EX_7}. So this power law is not able to describe the steep relaxation between $\lambda$=1 and $\lambda$=2 loop accurately. Microscopically the monotonic decrease of EB field with different field cycling can be explained with the help of two different uncompensated spins (frozen and rotatable) present at a magnetically disordered interface, affected by the magnetization reversals. These spins will contribute distinctively to the TE through different relaxation rates. Therefore we have used the following model proposed by Mishra \textit{et al.} to describe the TE completely\cite{EX_8}.
\begin{equation}
	H_{EB} (\lambda)= H_{EB}(\infty)+A_f exp(\frac{-\lambda}{P_f})+A_i exp(\frac{-\lambda}{P_i})
\end{equation}
where A$_f$ and P$_f$ are related to the frozen spin configuration. A$_i$ and P$_i$ are associated with flipping of rotatable spin. The A parameters have the dimension of magnetic field whereas the P parameters are dimensionless and resemble a relaxation time. It is seen that the Eq. (8) is fitted well to the experimental data including first two points than Eq. (7), dashed blue line represents the fitted curve shown in Figure 9(b). The estimated value of the fitted parameters are \textit{H}$_{EB}$ ($\infty$) = 1007 Oe $\pm$10 Oe which is close to the obtained value from Eq. (7), A$_f$= 148 Oe $\pm$3 Oe, P$_f$= 0.32 $\pm$0.15, A$_i$= 802 Oe $\pm$ 30 Oe, P$_i$ = 8.5 $\pm$ 0.28. 
The obtained value of $P_i$ is much larger than the value of $P_f$ and their ratio indicates the relative rate of their relaxation. It indicates that frustrated or rotatable uncompensated interfacial spins are contributing more to the TE and the frozen one relaxes slowly to the first one. 


\section{Conclusion}
In summary, the influence of ASD on the physical and magnetic properties of Gd-substituted YCMO has been investigated. Structural refinement confirms that the dominant phase of YGdCMO compound is monoclinic symmetry with space group P2$_1$/n. XPS analysis reveals that Co and Mn ions are present in mixed-valence states with Gd doping. Various experimental evidences confirmed that the multi-magnetic phases appear in the YGdCMO system. Multimagnetic orderings are achieved by Gd doping. This result is corroborated by our DFT calculations. AC susceptibility and memory effect measurements depict that the clusters do not form electronic-phase separation or spin-glass states. Although the FM transition temperature is observed at 95.5 K, the EB is visible only below \textit{T}$_N$=47 K, indicating that there is a connection between ASD and EB. The pinning of the magnetic moments at the interface of FM clusters separated by the AFM antiphase boundary is responsible for the observed EB effect with a large \textit{H}$_{EB}$= 1.07 kOe and H$_C$= 8.85 kOe. Due to various magnetic ordering and large EB fields, the sample can be considered as a potential candidate in spintronics applications. Therefore, it could be concluded that the EB effect and all the interesting properties originated from ASD present in the frustrated YGdCMO sample.   
\section{Acknowledgment}
The authors acknowledge Central Research Facility (CRF) and use of XPS under the DST-FIST facility in department of physics, IIT Kharagpur for this work. The authors also acknowledge use of SQUID in UGC-DAE, Indore for measurement. AK would like to acknowledge University Grants Commission (UGC) and Ministry of Education (MoE) for their financial support. VKM would like to acknowledge SMIL-13 grant from IIT Roorkee.


\end{document}